\documentclass[10pt,a4paper]{mjcs}
\usepackage{graphicx}
\usepackage {multirow}
\usepackage{float}
\usepackage[figurename=Fig.]{caption}
\newenvironment{mjcsabstract}{%
\begin{flushleft}
\textbf{\textit{ABSTRACT}}
\end{flushleft}

\it}

\title{Weighted Edge Sampling for Static Graphs}

\author{
		Muhammad Irfan Yousuf\thanks{Corresponding author: E-mail: dr.yousuf.irfan@gmail.com } \\
	Korea Institute of Science and Technology\\
	Seoul, South Korea\\
	\texttt{dr.yousuf.irfan@gmail.com}
	\and
	Raheel Anwar\\
	University of Graz\\
	Graz, Austria\\
	\texttt{raheel.anwar@alumni.uni-graz.at}
}

\usepackage{fancyhdr}

\begin{document}


\baselineskip12pt


\maketitle


\begin{mjcsabstract}
Graph Sampling provides an efficient yet inexpensive solution for analyzing large graphs. While extracting small representative subgraphs from large graphs, the challenge is to capture the properties of the original graph. Several sampling algorithms have been proposed in previous studies, but they lack in extracting good samples. In this paper, we propose a new sampling method called Weighted Edge Sampling. In this method, we give equal weight to all the edges in the beginning. During the sampling process, we sample an edge with the probability proportional to its weight. When an edge is sampled, we increase the weight of its neighboring edges and this increases their probability to be sampled. Our method extracts the neighborhood of a sampled edge more efficiently than previous approaches. We evaluate the efficacy of our sampling approach empirically using several real-world data sets and compare it with some of the previous approaches. We find that our method produces samples that better match the original graphs. We also calculate the Root Mean Square Error and Kolmogorov–Smirnov distance to compare the results quantitatively.
\end{mjcsabstract}


\section{Introduction}
In the last few years, there has been an explosive growth of online social networks (OSNs) that have attracted a lot of attention from all over the world including researchers. The popularity of online social networks e.g., Facebook and Twitter offered a great opportunity to the research community to develop new methods to process these massive graphs that can grow to millions or even billions of users. The huge user base of these networks provides an open platform for social network analysis including  topological properties of these networks \cite{Mislove,Ahn,Kumar}, analyzing user behavior \cite{Benevenuto}, social interaction characterization \cite{Wilson}, and information dissemination studies \cite{Kwak}.

However, the expensive processing of massive social network graphs hinders researchers from a better understanding of these graphs because even with well-equipped computers it requires a huge amount of time and computation overhead. Of the possible solutions, graph sampling provides an efficient and inexpensive solution. In graph sampling, we select a representative subgraph of the original graph such that the small subgraph retains the characteristics of the huge original graph. In other words, we scale-down the original graph while keeping its properties.

Many sampling algorithms, with their own pros and cons, have been proposed in the last decade or so. Node Sampling and Edge Sampling are two classical approaches to sample a graph. In node sampling, we first sample the nodes and then induce edges between them whereas in edge sampling we first sample the edges \cite{Nasreen}. However, in most real applications, we can not perform node or edge sampling directly due to all kinds of constraints, e.g. can not enumerate the ID space or the unavailability of the whole graph in advance. In such scenarios, traversal based sampling becomes a better choice because it explores the network as it proceeds. There are a number of traversal based approaches e.g., Breadth First Sampling \cite{Mislove}, List Sampling \cite{LS}, Random Walks \cite{RW} and Snow-Ball Sampling \cite{Survey1} etc. Nonetheless, Node Sampling and Edge Sampling can sample static graphs efficiently and could be equally applied to streaming graphs \cite{Nasreen2}.

In the most basic form of Edge Sampling (ES), we pick an edge uniformly at random from the list of edges. Given the fact that a real-world graph could have million of edges, picking edges at random could produce disconnected samples or the clustering coefficient of the graph could compromise along with its degree distribution. This problem was solved to some extent in Totally Induced Edge Sampling(TIES) \cite{Nasreen} where the authors induced all the edges between the sampled nodes instead of just the sampled edges in the sample graphs. In this paper, we extend Edge Sampling by introducing weights and thereby changing the probability with which an edge is sampled. In our proposed method, when an edge is sampled the weight of each of its neighboring edges increases and these edges are sampled with higher probability than others. This way, it is more likely to extract a sample with more connected neighborhood that ES or TIES and produce better samples than these methods. We compare our method with the mentioned methods and show that our approach produces better samples than the previous approaches.

The rest of the paper is organized as follows. We present our sampling method in detail in section 2. We present the evaluation criteria for sampling methods and experimental results in section 3. We review the previous sampling approaches briefly in section 4 and conclude the paper in section 5.

\section{Our Approach to Sampling}
In this section, we start with preliminaries and then present our sampling method named Weighted Edge Sampling (WES) and its variation Totally Induced Weighted Edge Sampling (TIWES).
\subsection{Preliminaries}
Given a big graph G = (V, E), where V = $\{v_1,v_2,v_3, ... , v_n\}$ is the set of nodes and E = $ \{e_1, e_2, e_3, ... , e_m\}$ is the set of edges, we extract a sample graph $G_s = (V_s, E_s)$ from G such that $V_s \subset V$ and $E_s \subset E$.  The resulting sample graph $G_s$ has $|V_s|$ number of vertices and $|E_s|$ number of edges in it. We consider undirected graphs in this work and represent an edge between nodes $v_i$ and $v_j$ as a tuple $e(v_i,v_j)$ where $v_i$ and $v_j$ are called the end nodes of the edge. Given a sampling fraction $\phi$ such that $|V_s|/|V|$ = $\phi$, the aim of sampling is to produce a sample with different values of $\phi$.

\subsection{Weighted Edge Sampling}
\begin{figure}[t]
	\centering
	\includegraphics[width = 160mm, height=60mm]{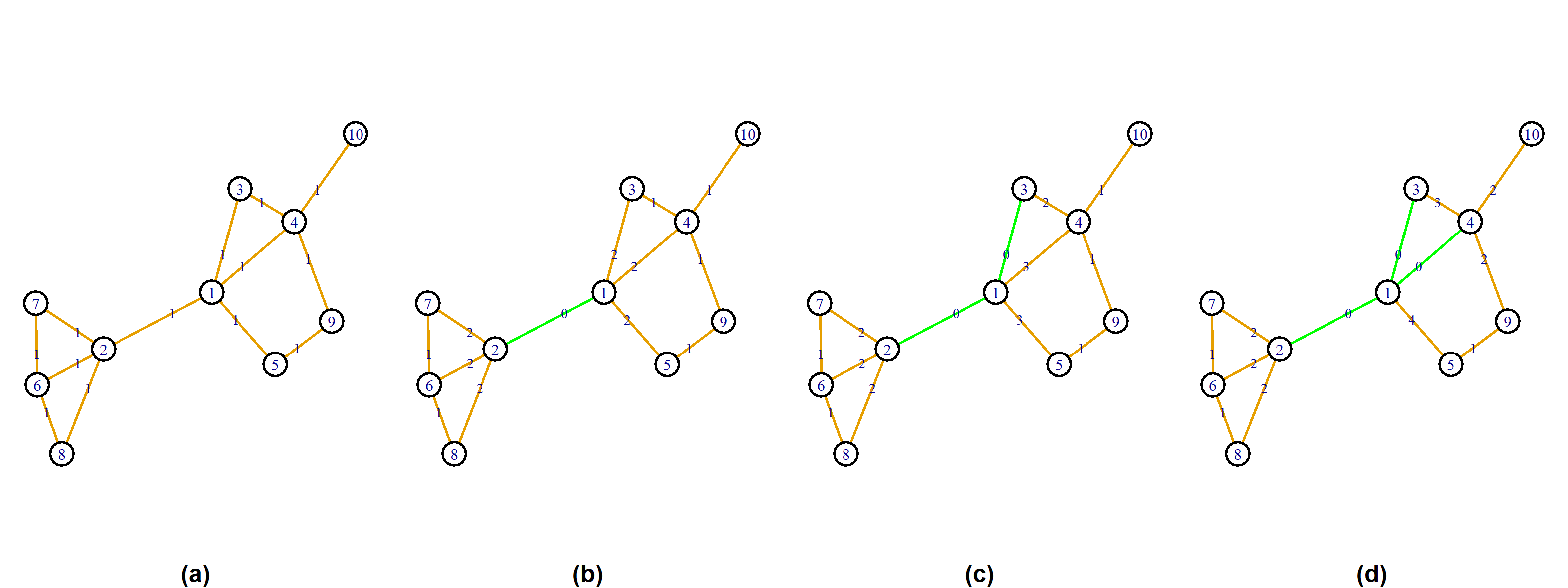}
	\caption{Working of Weighted Edge Sampling (a) The initial network with edge weight $w$=1 for all edges. (b) After edge e(1,2) is sampled, the weight of its neighboring edges increases (c) After edge e(1,3) is sampled. (d) After sampling edge e(1,4)}
	\label{fig_Exp}
\end{figure}

In Weighted Edge Sampling(WES), initially, all the edges in the graph hava a weight $w$=1. When an edge is sampled, the weight of its neighboring edges is increased by one. Let's consider an edge $e(v_i,v_j)$ between nodes $v_i$ and $v_j$ and let $w_e$ be the weight of this edge. The probability $p_e$ to sample this edge is given by
\begin{equation}
\ p_e=\frac{w_e}{|E|}
\end{equation}
where $|E|$ is the total number of edges. At start,  $w$=1 for all edges so an edge is sampled with uniform probability. When edge $e(v_i,v_j)$ is sampled, the weight of all of its neighboring edges is increased by one which increases their probability to be sampled. By neighboring edges we mean all the edges that have either $v_i$ or $v_j$ as an end node. We elaborate the method in figure \ref{fig_Exp}. Figure \ref{fig_Exp}(a) shows the original graph to be sampled with edge weights. Figure \ref{fig_Exp}(b) shows the graph when edge $e(1,2)$ has been sampled. We set the weight of sampled edges to zero while the weight of its neighboring edges is increased by one as shown in figure \ref{fig_Exp}(b), (c) and (d) as the sampling proceeds in WES. At the end of this sampling example in figure \ref{fig_Exp}(d), we have  $V_s = \{1, 2, 3, 4\}$ and  $E_s = \{e(1,2), e(1,3), e(1,4)\}$.

We believe that by increasing the weights and hence the probability of selecting the neighboring edges of a sampled edge, WES samples the neighborhood efficiently and produces good sampled subgraphs.

\subsection{Totally Induced Weighted Edge Sampling}
In most of sampled subgraphs, degrees of nodes are usually underestimated since only a fraction of neighbors may be selected. This results in a downward bias, regardless of what sampling algorithm is used as observed in \cite{Nasreen}. WES rectifies this effect to some extent, however, since the weight of neighboring edges is increased by one unit, therefore, WES can still miss some neighboring edges. To overcome this short coming, we apply graph induction \cite{Nasreen, LS} to WES and call the resulting method Totally Induced Weighted Edge Sampling (TIWES).

In TIWES, we induce all the edges between nodes in $V_s$ i.e., we add all the edges to $E_s$ between sampled nodes that are present in the original graph. For example, in figure \ref{fig_Exp}(d), the sampled node and edge sets would be $V_s = \{1, 2, 3, 4\}$ and  $E_s = \{e(1,2), e(1,3), e(1,4), e(3,4)\}$ in TIWES. We add $e(3,4)$ to $E_s$ because both of its end nodes are in $V_s$. With the induction step, TIWES can recover much of the connectivity of the original graph.

\section{Experimental Evaluation}
In this section, we evaluate the efficacy of both of our sampling algorithms, WES and TIWES, on eight real data sets and compare them with classic Edge Sampling (ES) and Totally Induced Edge Sampling (TIES) presented in \cite{Nasreen}.

\subsection{Data Sets}
In our experiments, we consider eight real networks including friendship networks, social networks, citation network and user-user interaction network. The size of these data sets range from 50,000 to more than 1 million nodes and edges between 200,000 to about 3 million. The main characteristics of these data sets are shown in table \ref{tab1}. These data sets are publicly available at \cite{konect}

\begin{table*}[!t]
	\small
	\caption {Real-world datasets used in the experiments}
	\begin{center}
		\begin{tabular}{ |p{1.5cm}|p{1.3cm}|p{1.3cm}|p{1.2cm}|p{1.2cm}|p{1.2cm}|p{5.6cm}|} \cline{1-7}
			\textbf{Datasets} &\textbf{Total Nodes} &\textbf{Total Edges} &\textbf{Average Degree} &\textbf{Average Clust. Coeff.} &\textbf{Average Path Length} &\textbf{Description} \\ \hline
			Brightkite &58,228 &214,078 &7.35 &0.27 &4.91 &Location-based social network\\			
			Facebook   &63,731 &817,035 &4.56 &0.02 &4.59  &User friendship network on Facebook\\
			LiveMocha  &104,103 &2,193,083 &6.84 &0.12 &4.90 &An online language learning community\\
			Gowalla    &196,591 &950,327 &9.67 &0.23 &4.62 &User-user friendship network\\
			DBLP       &317,080 &1,049,866 &6.62 &0.63 &6.75 &Citation network \\
			Amazon     &334,863 &925,872 &5.53 &0.39 &11.73 &Co-purchase of Amazon\\
			Twitter    &1,112,702 &2,278,852 &4.10 &0.02 &5.64 &Social network of twitter\\			
			YouTube    &1,134,890 &2,987,624 &5.27 &0.08 &5.55 &Friendship network of Youtube\\
			\hline
		\end{tabular}
	\end{center}
	\label{tab1}
\end{table*}

\subsection{Evaluation Criteria}
We evaluate the sampling algorithms primarily along three main properties; degree, clustering coefficient and path length. We measure the performance of a sampling algorithm by how well the sampled subgraphs preserve the average values and Cumulative Distribution Functions (CDFs) of each of these properties. We use five sampling fractions $\phi = \{0.02, 0.04, 0.06, 0.08, 0.1\}$. We show the following two measures.\\
\textbf{Point Statistics:} We calculate the ratio of the values of three properties of sampled subgraphs to the original values at all sampling fractions.\\
\textbf{Distributions} A distribution is a multivalued statistic and shows the distribution of a property in a graph. We measure and show the Cummulative Distribution Functions (CDFs) of degree, clustering coefficient and path length of sampled subgraphs at $\phi=0.06$ only along with the original distributions.\\
In addition to visually comparing the point statistics and distributions of the sampled subgraphs with those of the original networks, we also compute two statistics to compare them quantitatively. We use Root Mean Square Error (RMSE) for point statistics and  Kolmogorov–Smirnov Distance for distributions.\\
\textbf{Root Mean Square Error:} Given the original graph G and sampled graph $G_s$, we want to measure how far is $G_s$ from G. We use Root Mean Square Error (RMSE), given as
\begin{equation}
\, RMSE = \sqrt{\frac{1}{n}\sum_{1}^{n} (\Theta - \Theta_s)^2}
\end{equation}
where $\Theta$ and $\Theta_s$ are original and sampled values respectively.\\
\textbf{Kolmogorov–Smirnov Distance:} For distributions of the properties, we measure Kolmogorov–Smirnov (KS) distance. The Kolmogorov–Smirnov distance quantifies a distance between the empirical distribution functions of the sample and original distributions and gives the maximum distance between two distributions. It is calculated as
\begin{equation}
\, D_{ks} = \sup_{x} |(F_{n}(x)-F(x))|
\end{equation}
where $sup_x$ is the supremum of the set of distances.

\subsection{Results}
\begin{figure}[H]
	\centering
	\includegraphics[width = 160mm, height=240mm]{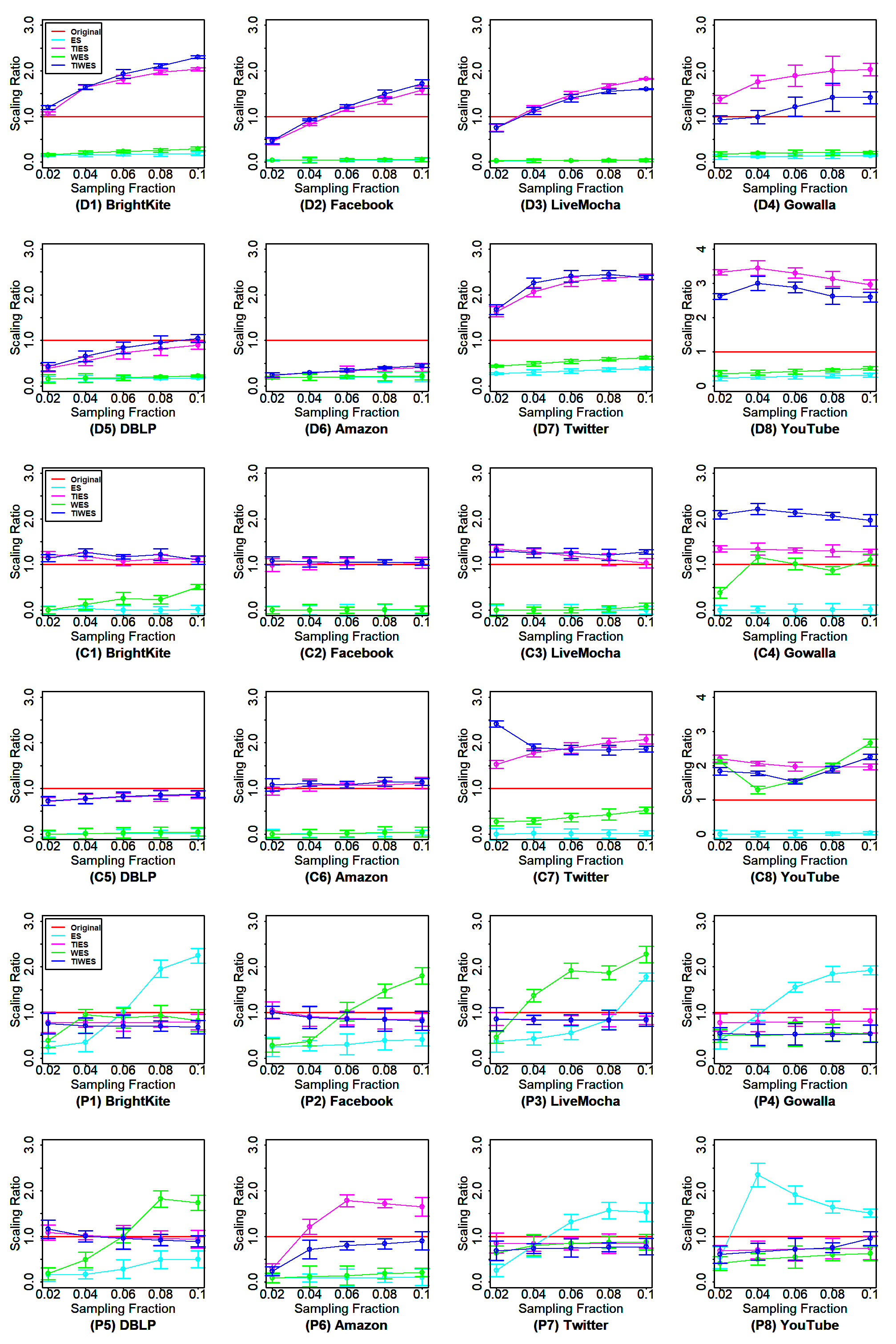}
	\caption{Point Statistics of all networks at different sampling fractions with 95\% confidence intervals. (D1-D8) Degree ,   (C1-C8) Clustering Coefficient   , (P1-P8) Path Length }
	\label{fig_PS}
\end{figure}

In our experiments, we obtain samples between 2 to 10\%. We use very small sampling fractions because the real-world graphs are huge and extracting a subgraph of 20\% (or higher) will be too big to serve the purpose of sampling. All the results presented in this section are averaged over five readings.\\
\textbf{Point Statistics: } We show the point statistics of average degree, clustering coefficient and path length for all data sets with 95\% confidence intervals in figure \ref{fig_PS}.  In figures \ref{fig_PS}(D1 to D8), we show the point statistics of average degree. We see that ES and WES perform poorly because these algorithms do not perform induction step and hence miss lots of connections. TIES and TIWES, on the other hand, induce all the edges between sampled nodes and perform better than other two methods. In datasets such as Gowalla and Youtube TIWES outperforms TIES while in other data sets both are comparable. In figures \ref{fig_PS}(C1 to C8), we show the results of clustering coefficient. Again, TIES and TIWES give better results than ES and WES. The performance of TIES and TIWES is comparable, TIES performing slightly better though. In figures \ref{fig_PS}(P1 to P8), we show the values of path length. We see that TIWES, on average, matches well with the original networks while TIES stands second to it. These results show that the sampled subgraphs produced by TIWES match well with the original graphs.\\
\textbf{Root Mean Square Error: }We compute the value of RMSE for each data set for degree, clustering coefficient and path length. We present the results in table \ref{tab2}. On average, TIWES generates small errors for degree and path length properties. Whereas, TIES performs the best for clustering coefficient property.\\
 
\begin{table*}[htb]
	\small
	\caption {RMSE values of  Degree, Clustering Coefficient and Path Length}
	\begin{center}	
		\begin{tabular}{|p{1.65cm}|p{0.7cm}|p{0.7cm}|p{0.7cm}|p{0.9cm}|p{0.7cm}|p{0.7cm}|p{0.7cm}|p{0.9cm}|p{0.7cm}|p{0.7cm}|p{0.7cm}|p{0.9cm}|} \cline{1-13}
			\multirow{2}{*}{\textbf{Data Set}} &\multicolumn{4}{c|}{\textbf{Degree}}  &\multicolumn{4}{c|}{\textbf{Clustering Coefficient}} &\multicolumn{4}{c|}{\textbf{Path Length}}  \\ \cline{2-13}
			&{ES}&{TIES}&{WES}&{TIWES}&{ES}&{TIES}&{WES}&{TIWES}&{ES}&{TIES}&{WES}&{TIWES} \\ \hline
			Brightkite &6.13&5.78&5.64&6.75&0.27&0.04&0.22&0.05&4.09&1.08&1.45&1.44\\
			Facebook   &24.53&10.35&24.38&11.92&0.25&0.01&0.25&0.01&2.94&0.52&2.59&0.58\\
			LiveMocha  &40.89&22.64&40.61&17.86&0.06&0.01&0.06&0.02&1.80&0.53&2.74&0.51\\
			Gowalla    &8.43&8.18&7.74&2.72&0.24&0.07&0.07&0.26&3.09&0.94&2.22&2.19\\
			DBLP	   &5.52&2.47&5.37&2.04&0.63&0.13&0.62&0.12&4.68&0.37&4.42&0.64\\
			Amazon	   &4.47&3.73&4.40&3.63&0.39&0.03&0.39&0.05&10.60&7.59&9.95&4.51\\
			Twitter	   &2.74&4.85&1.92&5.19&0.02&0.02&0.01&0.02&2.88&0.87&1.19&1.47\\
			YouTube	   &3.79&11.81&2.99&9.25&0.08&0.08&0.08&0.07&4.67&1.57&2.60&1.59\\ \hline
			Average    &12.06&8.73&11.63&\textbf{7.42}&0.24&\textbf{0.05}&0.21&0.07&4.34&1.69&3.39&\textbf{1.62}\\ \hline
			
		\end{tabular}		
	\end{center}
	\label{tab2}
\end{table*}

\begin{table*}[htb]
	\small
	\caption {KS distance of  Degree, Clustering Coefficient and Path Length}
	\begin{center}	
		\begin{tabular}{|p{1.65cm}|p{0.7cm}|p{0.7cm}|p{0.7cm}|p{0.9cm}|p{0.7cm}|p{0.7cm}|p{0.7cm}|p{0.9cm}|p{0.7cm}|p{0.7cm}|p{0.7cm}|p{0.9cm}|} \cline{1-13}
			\multirow{2}{*}{\textbf{Data Set}} &\multicolumn{4}{c|}{\textbf{Degree}}  &\multicolumn{4}{c|}{\textbf{Clustering Coefficient}} &\multicolumn{4}{c|}{\textbf{Path Length}}  \\ \cline{2-13}
			&{ES}&{TIES}&{WES}&{TIWES}&{ES}&{TIES}&{WES}&{TIWES}&{ES}&{TIES}&{WES}&{TIWES} \\ \hline
			Brightkite &0.51&0.16&0.46&0.18&0.44&0.19&0.42&0.23&0.53&0.41&0.18&0.39\\
			Facebook   &0.84&0.06&0.76&0.05&0.77&0.05&0.77&0.07&0.85&0.18&0.43&0.23\\
			LiveMocha  &0.77&0.06&0.73&0.05&0.70&0.19&0.70&0.18&0.59&0.29&0.98&0.28\\
			Gowalla    &0.61&0.07&0.71&0.01&0.59&0.11&0.58&0.08&0.85&0.33&0.84&0.38\\
			DBLP	   &0.77&0.23&0.72&0.21&0.85&0.31&0.84&0.29&0.62&0.07&0.41&0.09\\
			Amazon	   &0.87&0.48&0.84&0.37&0.79&0.51&0.79&0.51&0.98&0.72&0.99&0.41\\
			Twitter	   &0.11&0.41&0.06&0.32&0.08&0.21&0.08&0.21&0.96&0.37&0.33&0.45\\
			YouTube	   &0.31&0.25&0.33&0.30&0.23&0.28&0.13&0.47&0.83&0.43&0.77&0.41\\ \hline
			Average    &0.60&0.21&0.58&\textbf{0.19}&0.56&\textbf{0.23}&0.54&0.26&0.78&0.35&0.62&\textbf{0.33}\\ \hline
			
		\end{tabular}		
	\end{center}
	\label{tab3}
\end{table*}

\begin{figure}[H]
	\centering
	\includegraphics[width = 160mm, height=240mm]{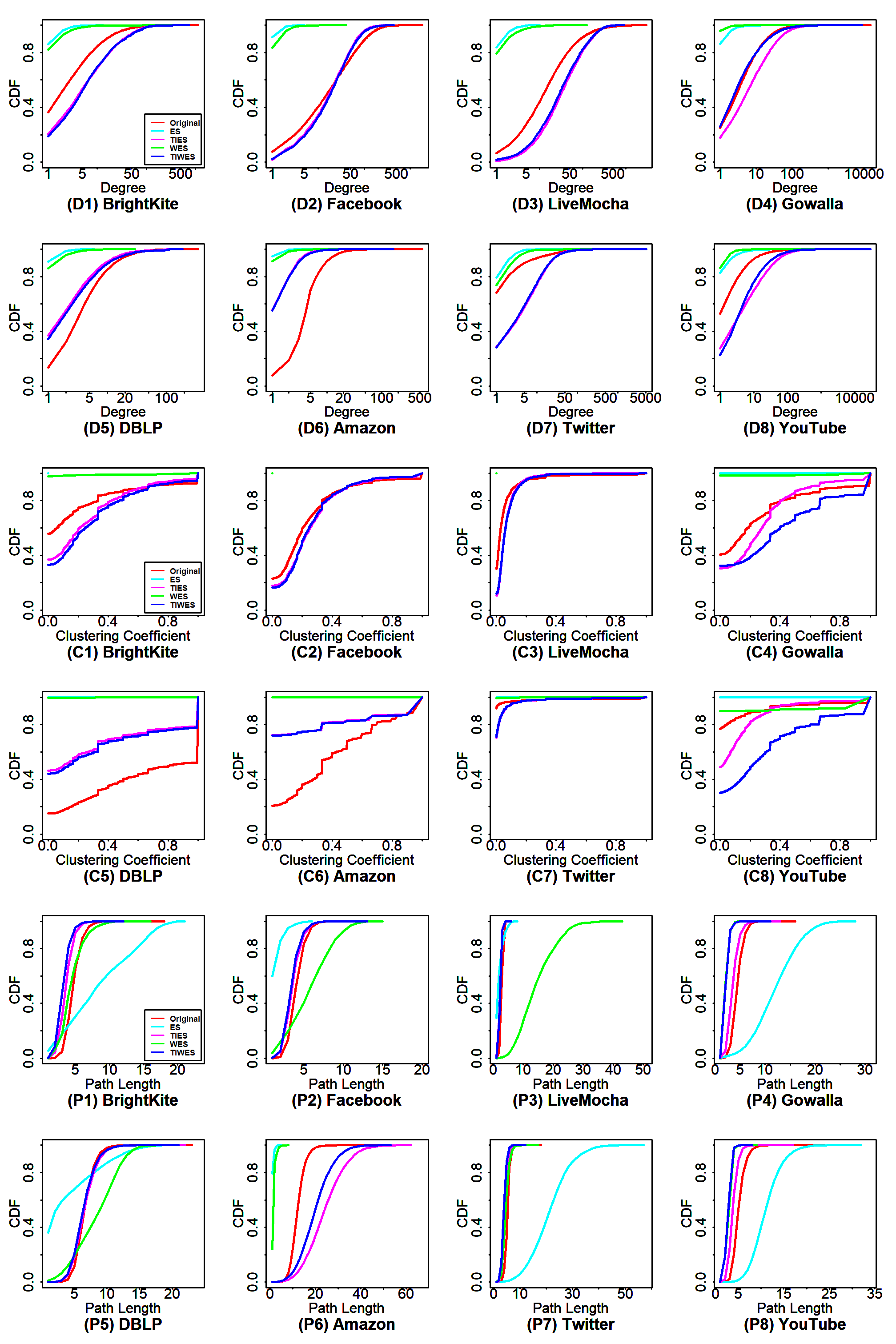}
	\caption{Distributions of all networks at $\phi=0.06$ (D1-D8) Degree ,   (C1-C8) Clustering Coefficient   , (P1-P8) Path Length}
	\label{fig_Dist}
\end{figure}

\textbf{Distributions: }We show the Cumulative Distribution Function (CDF) of degree, clustering coefficient and path length for all data sets at $\phi$=0.06 in figure \ref{fig_Dist}. The plots in figures \ref{fig_Dist}(D1 to D8) show the degree distributions and we see that, generally, ES and WES under-sample the degree while TIES and TIWES over-sample it. In case of clustering coefficient distributions, figures \ref{fig_Dist}(C1 to C8), both TIES and TIWES produce good samples. For path length distributions, we have mix results. In some data sets TIWES outperforms other methods while in other data sets TIES shows better results. Overall, it seems that graph induction plays an important role and combining it with edge weights can extract good samples from large graphs.\\
\textbf{Kolmogorov–Smirnov Distance: } We calculate Kolmogorov–Smirnov distance between the original distributions and sampled distributions at $\phi=0.06$ and present the results in table \ref{tab3}. The table shows that both TIES and TIWES tend to produce lower KS distance compared to ES and WES. Moreover, we see that TIES produces good samples in terms of clustering coefficient whereas TIWES generates samples that match the path length well with the original networks.

\section{Related Work}
Graph sampling has been used in many diverse fields of research including statistics, social science, data mining and machine learning, to name a few. The work in \cite{Rafiei,Gilbert} focuses on using sampling to reduce the graph for better visualization. The work in \cite{Lee, Yoon} studies the properties of samples of complex networks produced by traditional sampling algorithms such as node sampling, edge sampling and random walk based sampling. The authors in \cite{FFS} analyze various sampling algorithms for large graphs and propose Forest Fire Sampling (FFS). FFS is a partial breadth first sampling method in which we pick a seed node at random and then burn a fraction of its outgoing edges along with the nodes on the other end. Nasreen et al. proposed Totally Induced Edge Sampling (TIES) in \cite{Nasreen} which is a variation of Edge Sampling(ES). The primary difference between TIES and ES is the graph induction step. In TIES, we augment all the existing edges between the sampled nodes by including other edges between the set of sampled nodes in addition to those sampled in the edge sampling step. The authors in \cite{Doerr} discussed Breadth First Sampling (BFS), Depth First Sampling (DFS) and Random First Sampling (RFS) with focus on finding the minimum crawl size for estimating the properties of a graph. The List Sampling method \cite{LS} introduces the concept of a candidate list for keeping record of visited but yet unsampled nodes. With this list, authors propose a framework of different sampling methods with two parameters; one offers the flexibility of controlling the number of nodes to be sampled at a time while the second parameter is related to the probability with which a node is sampled. Random Walk is a classical random sampling method \cite{Xu1, Xu2}, which is a method of traversing a graph from node to node by randomly selecting neighbor nodes.The work in \cite{Maiya} provides a detailed study on the nature of biases in network sampling e.g., BFS usually produces biased samples with high degree nodes. Such biases can be rectified with Random Walks \cite{Lee}. The interested readers are referred to \cite{Survey1, Survey2} where the authors have surveyed different sampling approaches.

\section{Conclusion and Future Work}
We presented a new sampling method to extract representative subgraphs from large static graphs. In this sampling method, we give weights to the edges and edges are sampled with a probability proportional to their weights. When an edge is sampled, the weight of its neighboring edges increases by one and the probability of sampling these edges increases. This way our method explores the neighborhood of a sampled edge with higher probability. We also apply graph induction step that helps in retaining the connectivity in subgraphs. We apply our method on different real-world networks and compare with some of the existing approaches. The results show that our method produces better samples than previous approaches for certain properties of a graph. In future, we would like to extend our work to extract subgraphs from streaming and dynamic graphs. In addition, we would enhance it to work on directed graphs too.

\bibliographystyle{ieeetr}
\bibliography{File_bib}

\end{document}